\documentclass[11pt,twoside]{article}
\usepackage{asp2010}

\usepackage{graphicx}

\resetcounters

\begin{document}

\title{Gamma-ray emission from nova
outbursts}
\author{M. Hernanz
\affil{Institute of Space Sciences - ICE (CSIC-IEEC), Campus UAB, Fac. Ci\`encies, C5 par 2$^a$ pl.,
08193 Bellaterra (Barcelona), Spain}
}

\begin{abstract}

Classical novae produce radioactive nuclei which are emitters of gamma-rays in 
the MeV range. Some examples are the lines at 478 and 1275 keV (from $^7$Be and 
$^{22}$Na) and the positron-electron annihilation emission, with the 511 keV line 
and a continuum. Gamma-ray spectra and light curves are potential unique tools
to trace the corresponding isotopes and to give insights on the properties of 
the expanding envelope. Another possible origin of gamma-rays is the 
acceleration of particles up to very high energies, so that either neutral 
pions or inverse Compton processes produce gamma-rays of energies larger 
than 100 MeV. MeV photons during nova explosions have not been detected yet, 
although several attempts have been made in the last decades; on the other 
hand, GeV photons from novae have been detected with the Fermi satellite in 
V407 Cyg, a nova in a symbiotic binary, where the companion is a red giant with a wind, 
instead of a main sequence star as in the cataclysmic variables 
hosting classical novae. Two more novae have been detected recently (summer 
2012) by Fermi, apparently in non symbiotic binaries, thus challenging our 
understanding of the emission mechanism. Both scenarios (radioactivities and 
acceleration) of gamma-ray production in novae are discussed.
\end{abstract}

\section{Introduction}

The $\gamma$-rays are the most energetic photons in the Universe,  revealing very energetic phenomena, 
like the explosions of accreting white dwarfs in binary systems as classical or recurrent novae. 
Two types of $\gamma$-ray emission are expected from novae. First, $\gamma$-rays are emitted because 
radioactive nuclei are synthesized and ejected as a consequence of the explosion. Such emission occurs in the 
MeV energy domain (see Table \ref{tab:radioactivities}), and 
it traces nucleosynthesis directly. It has never been detected, neither with CGRO/Comptel \citep{Iyu95} nor with 
other previous satellites or the current INTEGRAL.
A second possible origin of $\gamma$-rays is related to particle acceleration in strong shocks between
nova ejecta and circumstellar material; this is particularly feasible when there's a red giant companion (i.e., in the 
symbiotic recurrent nova case), because nova ejecta shocks the red giant wind. But it could also occur whenever 
there is dense enough circumstellar matter to favor shocks. Two processes can explain the emission of VHE 
(very high energy) gamma-rays, with E larger than 100 MeV (i.e., in the GeV range): inverse Compton effect or 
neutral pion decay (see next section for details). This kind of emission has been detected already in three novae 
by the LAT instrument onboard the Fermi satellite \citep{che13}.

\section{``Fermi or GeV novae": Very High Energy (VHE) $\gamma$-rays (E $>$ 100 MeV)}
The first nova detected in VHE $\gamma$-rays was V407 Cyg \citep{abdo10}. This source is a binary system with a 
white dwarf and a Mira pulsating red giant companion. In March 2010 a nova outburst was detected from V407 Cyg. 
The Fermi/LAT telescope discovered a VHE $\gamma$-ray source (photons with E$>$100 MeV) which was coincident 
in position with V407 Cyg. As shown in Fig. 1 of \cite{abdo10}, such emission lasted for about two weeks 
after the nova eruption. It was not completely clear if the emission was originated by neutral pion decay or inverse Compton, 
the first corresponding to the hadronic scenario (accelerated protons are responsible) and the second one to the leptonic one 
(accelerated electrons are responsible). 
A detailed analysis presented in \cite{MD2013}, favored the leptonic scenario, where accelerated electrons upscatter 
(Inverse Compton) nova light up to very high energies.

Before V407 Cyg was detected by Fermi/LAT, a theoretical prediction of particle acceleration during the 2006 eruption 
of the recurrent symbiotic nova RS Oph was made by \cite{TH07}, as well as its ensuing VHE $\gamma$-ray emission 
\citep{HT12}. The previous eruption of this nova was in 1985, so the recurrence period is 21 years. A self consistent 
thermonuclear runaway model for this nova is not easy to find, because a combination of high accretion rate 
(at least $2\times 10^{-7}~M_\odot$~yr$^{-1}$) and extremely large initial mass of the white dwarf (at least $1.35~M_\odot$) 
is required \citep{HJ08}. When the nova explodes, an expanding shock wave sweeps the red giant wind, and the system behaves 
as a ``miniature" supernova remnant, much dimmer and evolving much faster. 

The evolution of the blast wave of RS Oph (2006 outburst) is shown in Figure 1 from \cite{TH07}, where  the time dependence of the 
forward shock velocity as deduced from IR spectroscopic observations is compared to that from the X-ray observations with RXTE. 
Two caveats are, first why the cooling phase started as early as 6 days, when T(shock) was about 
$10^8$ K and thus radiative cooling was not important, and second why shock velocities derived from X-rays are lower than those from IR 
measurements. The answer is that there was particle acceleration (i.e., generation of cosmic rays), with the ensuing energy loss 
(associated with particle escape). Such losses were much more efficient (more than 100 times larger) than radiative losses to cool the shock, 
explaining the very fast cooling. They also explain the lower shock velocity deduced from X-ray observations, because the usual relation 
for a test-particle strong shock underestimates the shock velocity when particle acceleration is efficient, because the shock temperature is 
lower (see \cite{TH07}). 

A prediction of the $\gamma$-ray emission associated to the accelerated particles was made \citep{HT12}. The production of neutral pions 
($\pi^0$) was calculated from the density in the red giant wind and the cosmic-ray energy density required to explain the IR and X-ray observations. 
The Inverse Compton (IC) contribution was estimated from the non thermal synchrotron luminosity (deduced from the early radio detections of RS Oph at 
frequencies below 1.4 GHz by \cite{Kanth07}). Pion decay dominates over inverse Compton in RS Oph, but this should not be the general rule 
for all novae in symbiotic binaries. Comparison between theoretical predictions for RS Oph and Fermi/LAT sensitivities shows that RS Oph would 
have been detected by Fermi/LAT, if it had been in orbit in 2006. 

In addition to V407 Cyg, two more novae - Nova Sco 2012 and Nova Mon 2012 - have also been detected by Fermi/LAT at E$>$100 MeV. However, the 
companions of these exploding white dwarfs are not red giants, so the scenario is different, although a dense circumstellar environment (``playing the role" 
of the  missing red giant wind) exists, at least for Nova Mon 2012, which in fact was first discovered in $\gamma$-rays than optically (see \cite{Sho13} for 
details about this nova). 

\section{$\gamma$-rays from radioactivities: E $\sim$ 1 MeV}
The potential role of novae as $\gamma$-ray emitters was already pointed out in the 70's \citep{CH74} 
and 80's \citep{Cla81}; see recent review in \cite{Her08}.
The $\gamma$-ray signatures of classical novae depend on their yields of 
radioactive nuclei. CO and ONe novae differ 
in their production of $^{7}$Be and $^{22}$Na, whereas they 
synthesize similar amounts of $^{13}$N and $^{18}$F. Thus CO novae should 
display line emission at 478 keV related to $^{7}$Be decay, whereas for ONe 
novae line emission at 1275 keV related to $^{22}$Na decay is expected. In 
both nova types, there should be as well line emission at 511 keV related to 
e$^-$--e$^+$ annihilation, and a continuum produced by Comptonized 511 keV 
emission and positronium decay. In Table \ref{tab:radioactivities} the main properties of 
the radioactive nuclei synthesized in novae are shown.

All the results presented in this paper correspond to the most recent nucleosynthetic yields (J. Jos\'e, unpublished) 
of nova models computed with the SHIVA code, described in \cite{JH98}, and the nuclear reaction rates from Iliadis et al.
The most significant change is the reduction of the amount of $^{18}$F, in the last years, 
because of revised rates of the nuclear reactions affecting its production and destruction (as reported in 
\cite{Her99,Coc00,Ser03,Cha05}, and others).

The shape an intensity of the $\gamma$-ray output of novae as well as its temporal 
evolution, depend on the amount of $\gamma$-ray photons produced and  
on how they propagate through the expanding envelope and ejecta \citep{LC87,Gom98}. 
Interaction processes affect the propagation of photons, i.e. Compton 
scattering, e$^-$--e$^+$ pairs production and photoelectric absorption. 

\begin{figure}
\centerline{
\hspace{2cm}
\includegraphics[width=8.5cm]{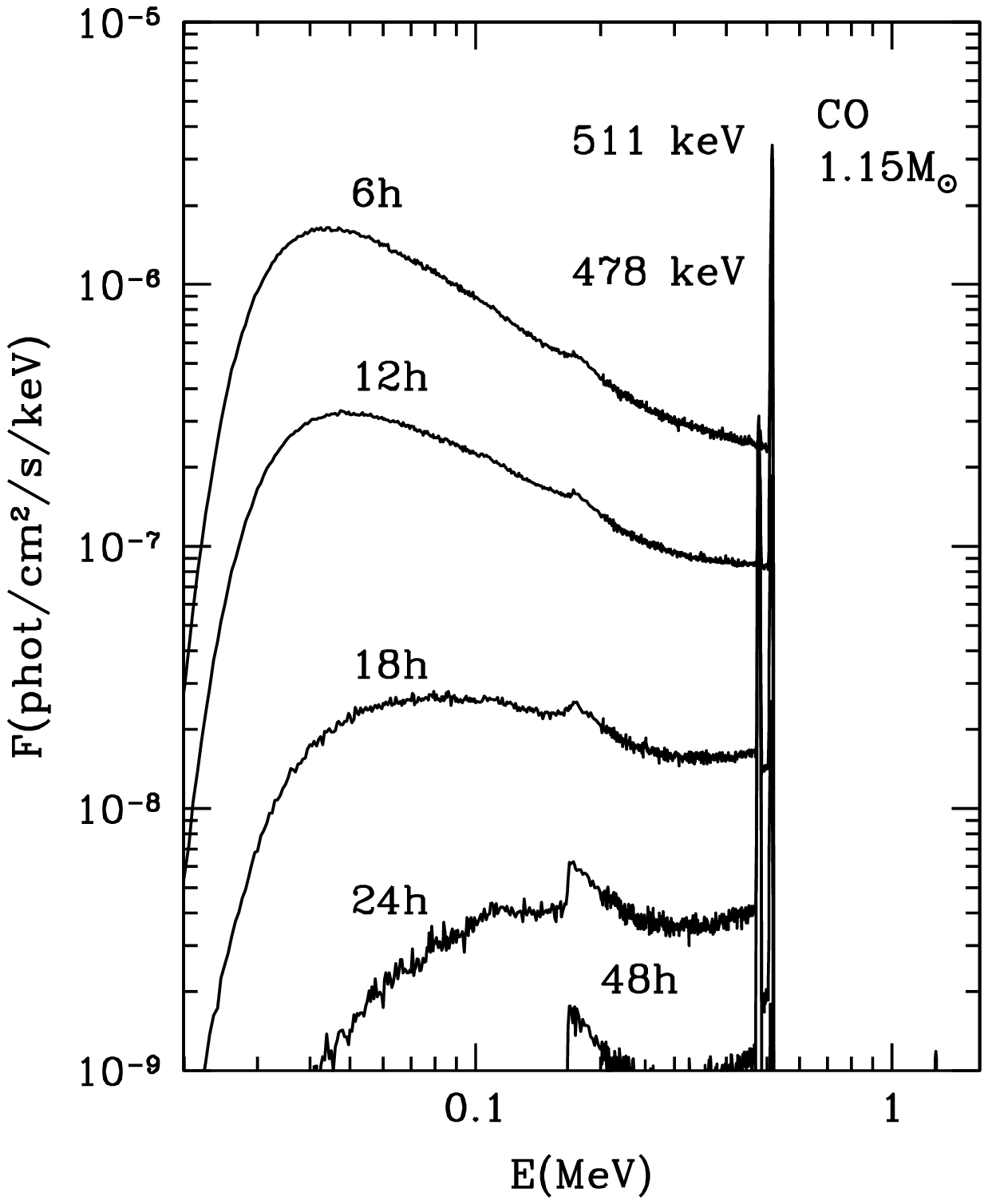}
\hspace{-3cm}
\includegraphics[width=8.5cm]{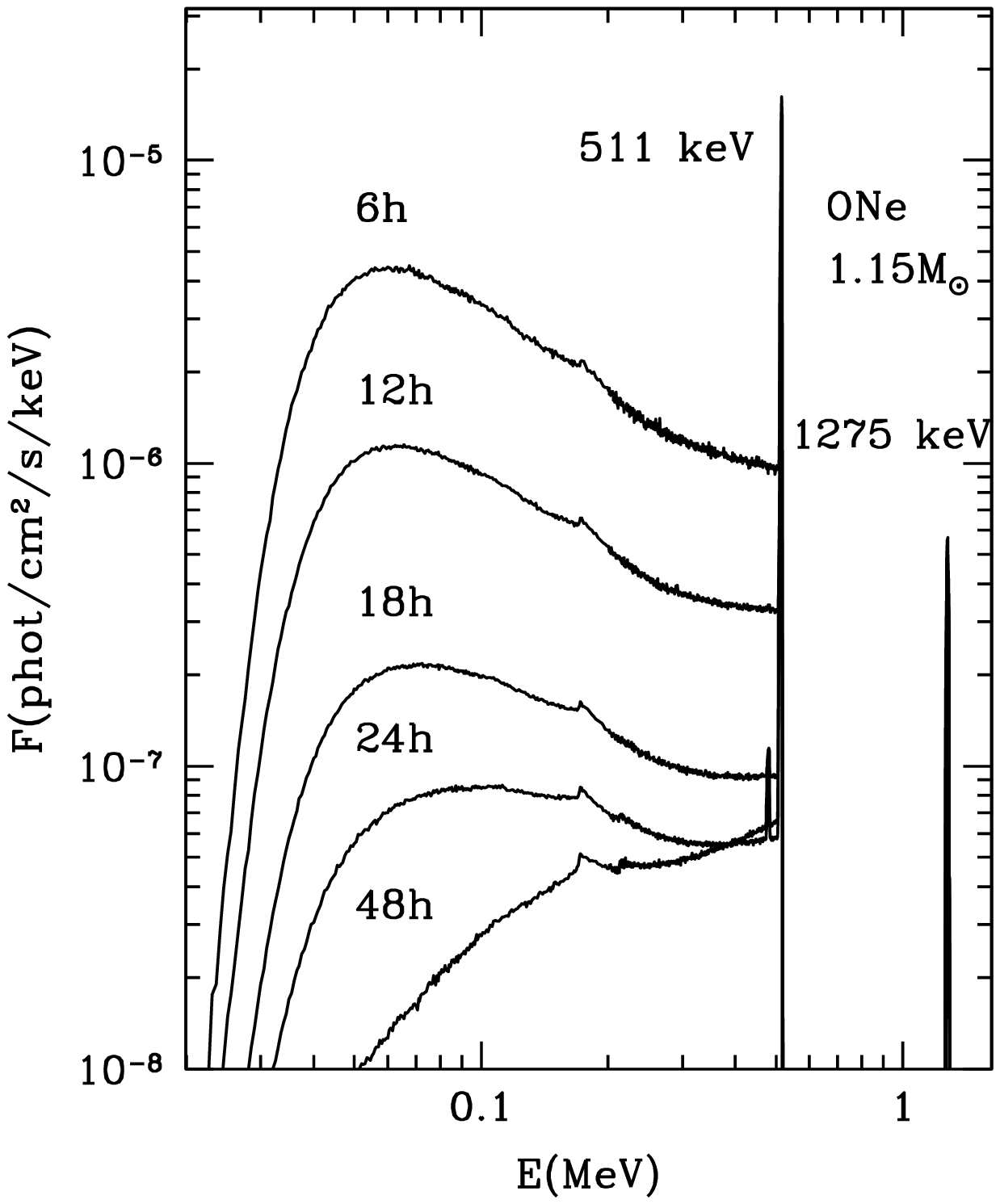}
}
\caption{Left panel: Spectra of a CO nova of mass 1.15 M$_\odot$ at different epochs after T$_{peak}$; 
Right panel: Same for an ONe novae of the same mass.}
\label{fig:specCO_ONe}
\end{figure}

A Monte Carlo code, based on the method described by \cite{PSS83} and \cite{AS88}, 
was developed by \cite{Gom98} to compute the $\gamma$-ray output of novae. 
The temporal evolution of the whole $\gamma$-ray spectrum of some  
representative models is shown in Figure \ref{fig:specCO_ONe}. 
The most prominent features are the annihilation line at 511 keV and 
the continuum at energies between 20-30 keV and 511 keV (in both nova 
types), the $^{7}$Be line at 478 keV in CO novae, and the $^{22}$Na line 
at 1275 keV in ONe novae. 

\begin{figure}
\centerline{
\hspace{2cm}
\includegraphics[width=8.5cm]{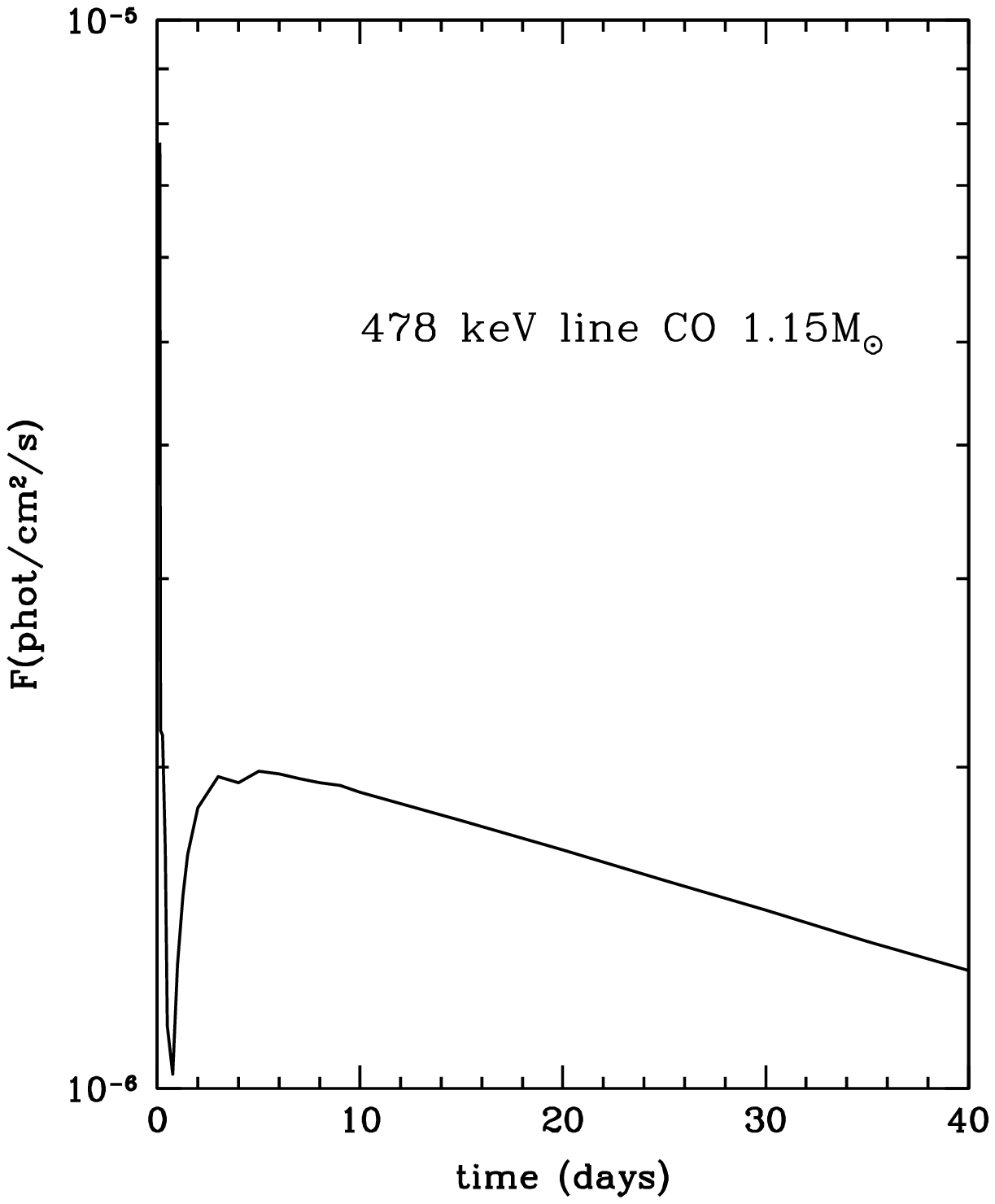}
\hspace{-3cm}
\includegraphics[width=8.5cm]{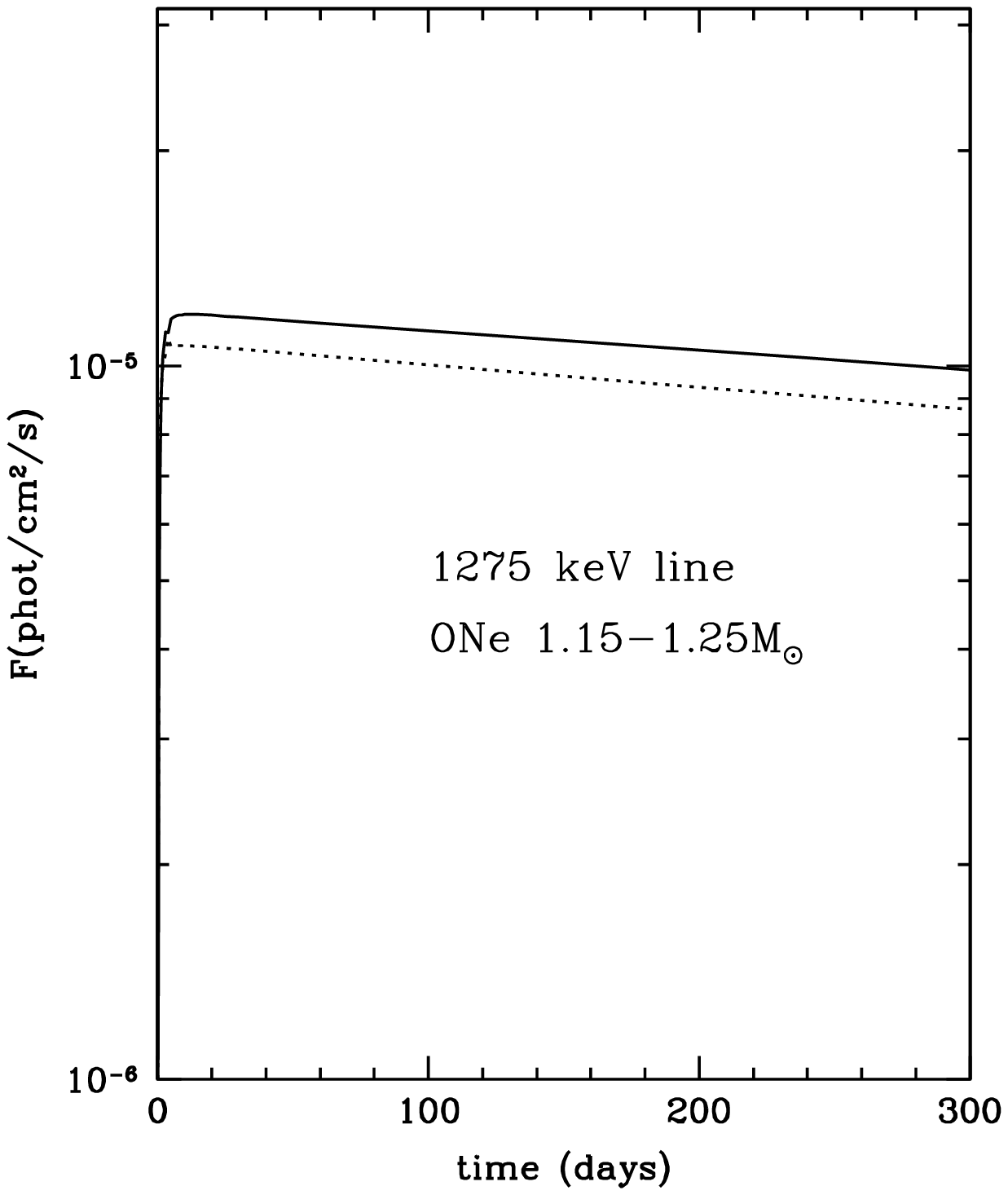}
}
\caption{Left panel: Light curve of the 478 keV line for a CO nova with mass 1.15 M$_\odot$. 
Right panel: Light curve of the 1275 keV line for ONe novae of 1.15 and 1.25 M$_\odot$ (solid and dotted, respectively).}
\label{fig:lcBe_Na}
\end{figure}

The light curves of the 478 keV line are shown in 
Figure \ref{fig:lcBe_Na}:  the flux reaches its maximum ($\sim10^{-6}$phot cm$^{-2}$ s$^{-1}$, for d=1kpc) 
at day $\sim$5 in the model with mass 1.15 M$_\odot$. 
The width of the line is $\sim$8 keV. There is a 
previous maximum, which has nothing to do with the envelope's content of $^{7}$Be, 
but with the strong continuum related to the annihilation of 
$^{13}$N and $^{18}$F positrons. 

The $^{22}$Na line at 1275 keV appears only in ONe novae, because CO novae do not 
synthesize this isotope. The rise phase of the 1275 keV line 
light curves lasts between 10 (1.25 M$_\odot$) and 20 days (1.15 M$_\odot$). 
After the maximum (flux $\sim 10^{-5}$phot cm$^{-2}$ s$^{-1}$ at d=1kpc), the 
line reaches the stable decline phase dictated by the lifetime of  $^{22}$Na, 
3.75 years; during this phase, the line intensities directly reflect the amount 
of $^{22}$Na ejected mass (see Figure \ref{fig:lcBe_Na}).  
The width of the line is around 20 keV, which is a problem for its  
detectability with instruments having good spectral resolution, which are best suited for narrow lines 
(e.g., Ge detectors of SPI on board INTEGRAL).

\begin{figure}
\centerline{
\hspace{2cm}
\includegraphics[width=8.5cm]{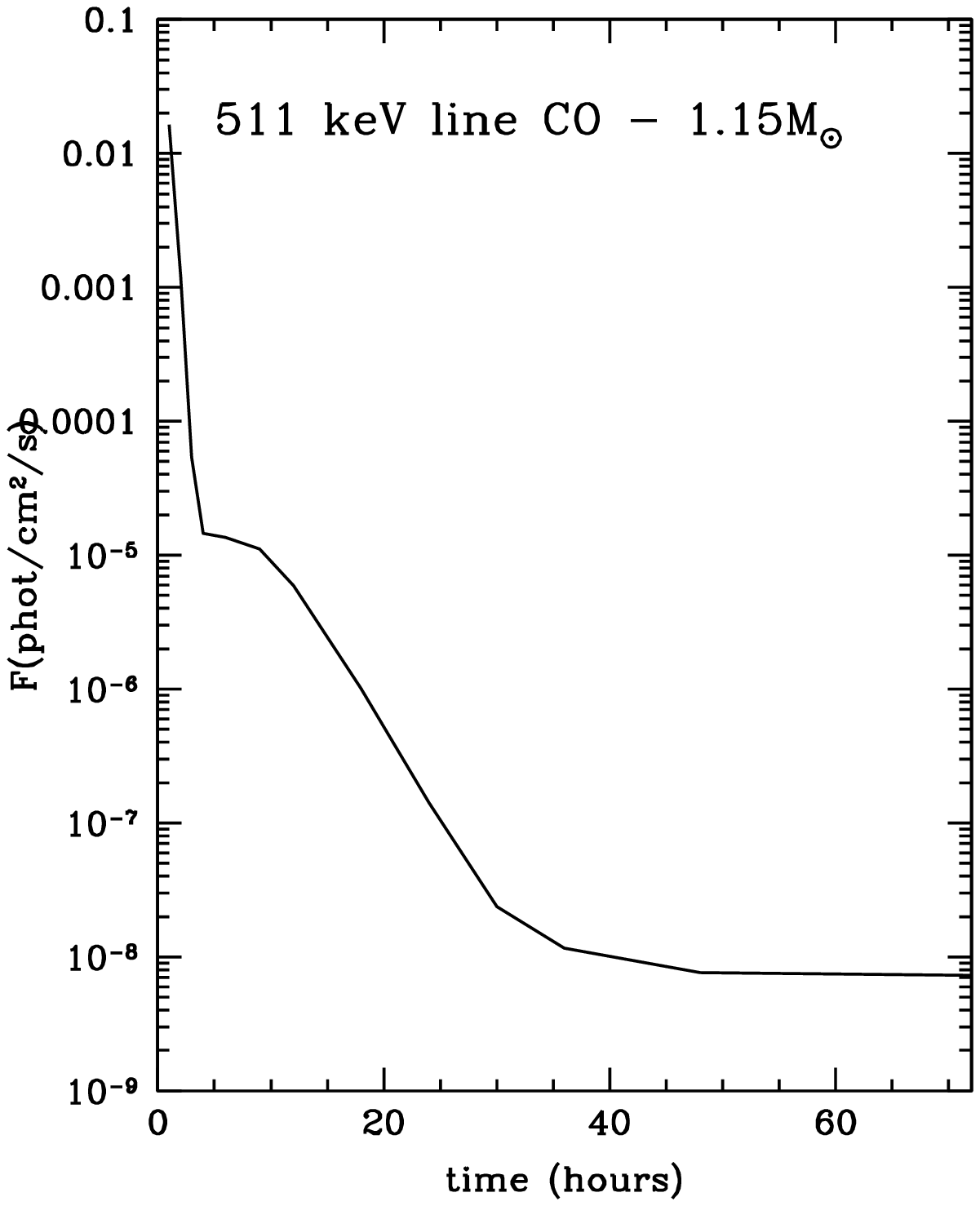}
\hspace{-3cm}
\includegraphics[width=8.5cm]{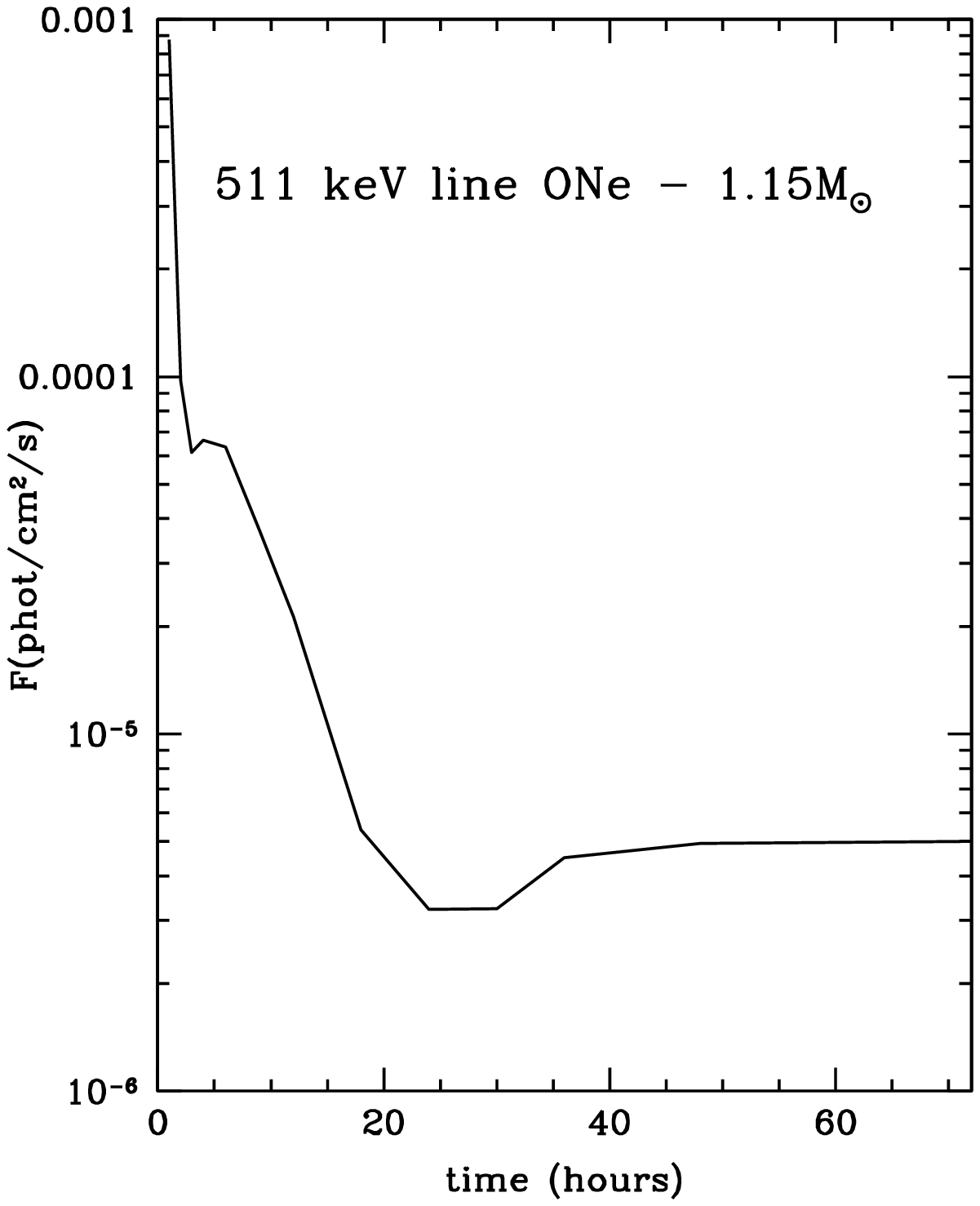}
}
\caption{Left panel: Light curve of the 511 keV line for a CO nova with mass 1.15 M$_\odot$. 
Right panel: Same for an ONe nova of 1.15 M$_\odot$.}
\label{fig:lc511}
\end{figure}

\begin{figure}
\centerline{
\hspace{2cm}
\includegraphics[width=8.5cm]{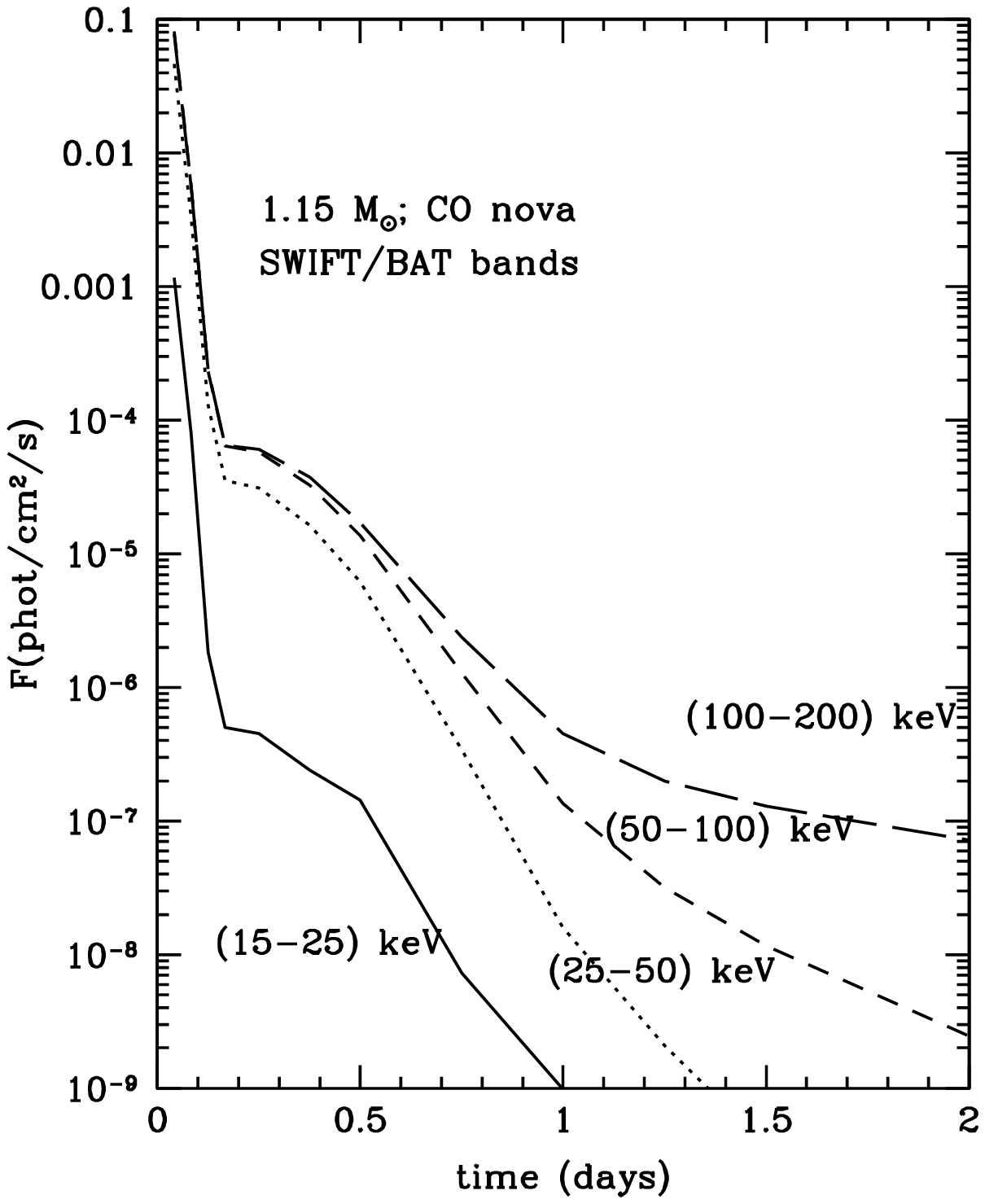}
\hspace{-3cm}
\includegraphics[width=8.5cm]{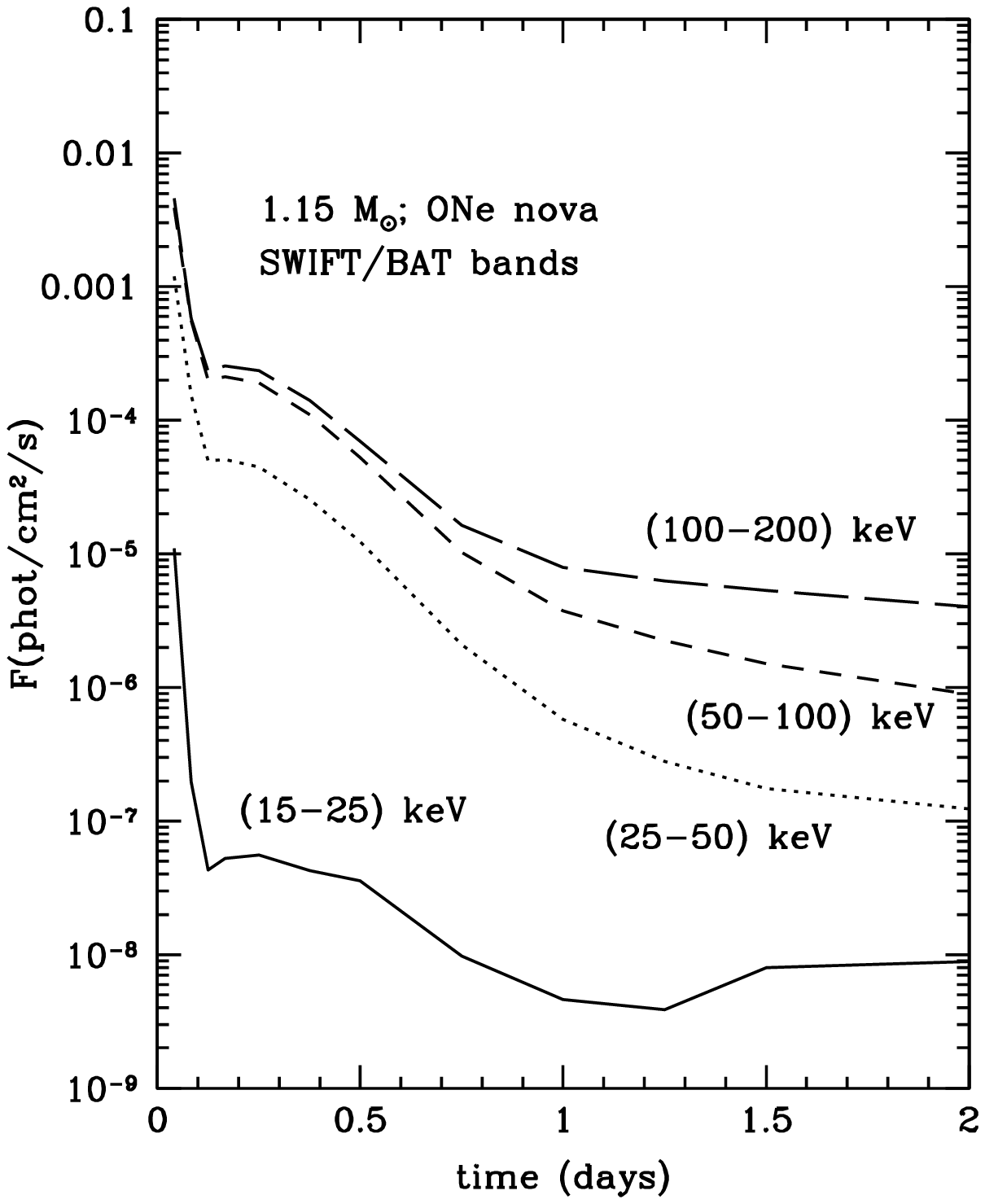}
}
\caption{Light curves for various continuum bands appropriate for the Swift/BAT instrument. Left panel: CO nova with mass 1.15 M$_\odot$. 
Right panel: Same for an ONe nova of 1.15 M$_\odot$.}
\label{fig:lcswift}
\end{figure}

The early $\gamma$-ray emission of novae is related 
to the disintegration of the very short-lived $\beta^+$-unstable 
isotopes $^{13}$N and $^{18}$F. The radiation is emitted as 
a line at 511 keV plus a continuum related with both the
positronium continuum and the Comptonization of the photons 
emitted in the line. The sharp cut-off at energies 20-30 keV is caused by 
photoelectric absorption. The light curves of the 511 keV line are shown in 
Figure \ref{fig:lc511} for a CO and an ONe novae. Larger fluxes are emitted in 
the continuum (e.g., in Figure  \ref{fig:lcswift} are shown 
the light curves for various continuum bands appropriate for Swift/BAT). 
The two maxima in the light curves correspond 
to $^{13}$N and $^{18}$F decays, but the first maximum is difficult to resolve 
because its duration is really short; in addition, it is very model dependent: 
only $^{13}$N in the outermost zones of the envelope can be seen in $\gamma$-rays 
because of limited transparency at very early epochs and, therefore, 
the intensity of the first maximum depends on the efficiency of convection. 
This first maximum gives thus an important insight into the dynamics of the envelope after 
peak temperature is attained at its base.

The annihilation emission is the most intense $\gamma$-ray feature expected from novae, 
but unfortunately it has a very short duration, because of the short lifetime 
of the main positron producers ($^{13}$N and $^{18}$F). There are also positrons 
available from $^{22}$Na decay in ONe novae, but these contribute much less 
(they are responsible for the {\it plateau} at a low level, between $10^{-6}$ and $10^{-5}$ 
phot cm$^{-2}$ s$^{-1}$ for d=1kpc; see Figure \ref{fig:lc511}). 
These positrons do not contribute all the time, because after roughly one week the envelope is so transparent 
that $^{22}$Na positrons escape freely without annihilating. In summary, annihilation 
radiation lasts only $\sim 1$ day at a high level, and one to two weeks at a lower level 
{\it plateau} (the latter mainly in ONe novae, but also at a very much lower level in CO ones, 
which have a $\sim1000$ times smaller amount of $^{22}$Na, ).

An important fact is that annihilation radiation is emitted well before the 
visual maximum of the nova, i.e. before the nova is discovered optically.
This early appearance of $\gamma$-rays from electron-positron annihilation makes their detection 
through pointed observations almost impossible. Only wide field of view instruments, 
monitoring continuously the sky in the appropriate energy range can detect it. 
A detailed study has been performed with the Burst Alert Telescope (BAT) onboard the Swift satellite 
\citep{Sen08}. Swift/BAT offers an interesting opportunity to search for annihilation emission, 
because of its huge field of view, good sensitivity, and well-suited energy band (14-200 keV). Data from Swift/BAT can be retrospectively 
analyzed to search for prompt $\gamma$-ray emission from the direction of
novae after their optical discovery. The search for emission from the 24 classical novae discovered since the Swift launch 
yielded no positive results, which was understood since none of them was close enough. Other previous searches were made with the TGRS 
instrument onboard the Wind satellite \citep{Har99}, and the BATSE instrument onboard the CGRO satellite \citep{Her00}.

\begin{table}[h]
\caption{Main radioactive nuclei synthesized in nova explosions.}
\label{tab:radioactivities}
\begin{tabular}{ccccc} 
\hline \hline
Isotope   & Lifetime       & Main disintegration  & Type of emission
          & Nova type\\
          &                            &  process                      & 
          &\\
\hline
$^{13}$N  & 862~s          & $\beta^+$-decay              
                                             & 511~keV line and
          & CO and ONe\\      
          &                &         &    continuum        & \\
$^{18}$F  & 158~min        &  $\beta^+$-decay      
                                              & 511~keV line and 
          & CO and ONe\\
          &                &         &   continuum         & \\
$^{7}$Be  & 77~days        & $e^-$-capture      
                                              & 478~keV line 
          & CO\\
$^{22}$Na & 3.75~years  &  $\beta^+$-decay
                                              & 1275 and 511~keV lines  
          & ONe\\
$^{26}$Al & 10$^{6}$~years &  $\beta^+$-decay
                                              & 1809 and 511~keV lines 
          & ONe\\
\hline \hline
\end{tabular}
\end{table}

\section{Summary and conclusions}
The long awaited detection of $\gamma$-rays from novae has been accomplished with the Fermi/LAT \citep{abdo10}; this instrument observed emission 
of VHE $\gamma$-rays, produced either by neutral pions or by inverse Compton, as a consequence of particle acceleration 
in the shock wave from the interaction between the nova ejecta and the red giant wind in V407 Cyg (or with other circumstellar material 
in N Sco 2012 and N Mon 2012) . RS Oph, a recurrent nova in a symbiotic binary, i.e., with a red giant companion, had already been predicted 
to emit such VHE photons \citep{TH07}.  

On the contrary, the detection of $\gamma$-rays in the MeV range, from nova radioactivities, has not been achieved yet. The predictions for the current INTEGRAL/SPI 
instrument are not very optimistic: distances shorter than 0.5 kpc for the $^7$Be line at 478 keV, and shorter than 1 kpc for the  $^{22}$Na line at 1275 keV, 
are required. A future generation of 
instruments is needed, either a powerful Compton Telescope (e.g., the ACT project) or a $\gamma$-ray lens - Gamma-Ray Imager, GRI - or a 
combination of both, DUAL (see e.g. \cite{boggs04} and \cite{pvb12}). The Laue $\gamma$-ray lens provides up to now the best perspectives for detecting lines in the MeV range, 
since a large collecting area (the diffracting crystals acting as a photon concentrator) is combined with a small detector in its focal plane, thus yielding a good signal to noise ratio, not easily reachable with Compton Telescopes. We should wait several years until this already proven concept is feasible for a space mission.

\acknowledgements The author thanks funding from the Spanish MINECO project AYA2011-24704, FEDER funds and the AGAUR (Generalitat of Catalonia) 
project 2009 SGR 315.


\end{document}